\documentstyle[aps,prb,multicol,epsfig]{revtex}
\newcommand \sech { \mathrm{sech}}
\renewcommand \Im {\mathrm{Im}}
\begin{document}
\title{Effects of anisotropic spin-exchange interactions in spin ladders}
\author{R. Citro$^{a}$ and E. Orignac$^{b}$}
\address{
$^{a}$Dipartimento di Fisica ''E.R. Caianiello'',\\
University of Salerno and Unit\`{a} INFM of Salerno, Baronissi (Sa), Italy\\
$^{b}$ Laboratoire de Physique Th\'eorique, CNRS UMR 8549, \\
 Ecole Normale   Sup\'erieure, 24 Rue Lhomond 75231 Paris Cedex 05, France \\}

\date{\today}
\maketitle

\begin{abstract}
We investigate the effects of the Dzialoshinskii-Moriya (DM) and
Kaplan-Shekhtman-Entin-Wohlman-Aharony (KSEA)
interactions on various thermodynamic and magnetic properties of a
 spin 1/2 ladder.  Using  the Majorana
fermion representation, we derive the spectrum
of low energy excitations for a pure DM
 interaction and in presence of a superimposed KSEA
interaction. We calculate the various correlation functions for both
cases and discuss how they are modified with respect to the case of an
isotropic ladder. We also discuss the electron spin resonance (ESR) spectrum of the system and show
that it is strongly influenced by the orientation of the magnetic
field with respect to the Dzialoshinskii-Moriya vector. Implications of our calculations
for NMR and ESR experiments on ladder systems are discussed.
\end{abstract}


\bigskip

\section{Introduction}
In recent years low dimensional quantum spin systems have attracted
great interest due to their fascinating properties originating from low
dimensionality and quantum fluctuations. Quantum effects are
particularly relevant in $S=1/2$ systems. The quasi-one-dimensional
antiferromagnetic $S=1/2$ ladders have been extensively investigated
both theoretically and
experimentally\cite{dagotto_2ch_review}. Several
experimental realizations of such systems are available\cite{azuma_srcuo,siegrist_srcacuo,chaboussant_cuhpcl}. In particular
${\rm Sr_{n-1}Cu_{n+1}O_{2n}}$ and ${\rm Sr_{14}Cu_{24}O_{41}}$ seem to be quite
well described by a simple isotropic Heisenberg ladder
models\cite{dagotto_2ch_review}. However, countless
experiments on higher dimensional compounds
\cite{ishikawa_ferro,lebech_ferro,zheludev_bacugeo,kataev,tsukada}
indicate that in many systems the isotropic exchange interaction is
not sufficient to describe magnetic properties such as weak ferromagnetism.
From the theoretical point of view, more than thirty years ago
Dzialoshinskii\cite{dzyalo_interaction} pointed out  that a term
(allowed by symmetry in non centrosymmetric structures) of
the form ${\mathbf{D}}\cdot ({\mathbf{S}}_1\times {\mathbf{S}}_2)$, where
${\mathbf{D}}$ is the constant Dzialoshinskii vector and
${\mathbf{S}}_{1,2}$ are the sublattice magnetizations,  favors a
canted spin arrangement over the antiferromagnetic one and a weak
ferromagnetic moment. This term was
later derived by
Moriya\cite{moryia_asym_int} from a microscopic Hubbard-type
Hamiltonian  by including   spin-orbit coupling in the
Anderson superexchange calculation\cite{anderson_superexch}. Actually,
the calculation of Moriya showed that besides the
Dzialoshinskii-Moriya (DM) term, a symmetric  anisotropy term
of the form ${\bf S}\cdot \tensor{A} \cdot {\bf S}$ was also obtained,
but was assumed by Moriya to be negligible compared to the
antisymmetric one.  However, it has been  realized recently
\cite{kaplan_anisotropic,shekhtman_anisotropic_s,shekhtman_anisotropic_l}
that this assumption is incorrect if the underlying microscopic model
from which spin-spin interactions are derived
has an SU(2) spin symmetry. In that case, the effective spin-spin
interactions have the same SU(2) invariance as the microscopic
model. As the DM term alone breaks SU(2) symmetry, the symmetric
 term has to be such that it compensates exactly the spin anisotropy
induced by the DM term. For this class of systems, the symmetric
anisotropy term is called the Kaplan-Shekhtman-Entin-Wohlman-Aharony
(KSEA) and it is tuned in such way that SU(2) symmetry is
recovered\cite{kaplan_anisotropic,shekhtman_anisotropic_s,shekhtman_anisotropic_l}.
A  derivation of the DM and the KSEA term starting from a
microscopic SU(2) invariant Hubbard model with spin-orbit coupling
can be found in  \cite{shekhtman_anisotropic_s}.  

The main objective of this paper is to study the influence of the
anisotropic (DM and KSEA) spin-exchange interactions in a spin 1/2 two leg
ladder. We will consider a
uniform  Dzialoshinskii-Moriya interaction in contrast
with the staggered interaction
discussed in \onlinecite{oshikawa_cu_benzoate}.  We mainly
concentrate on a DM interaction along the rung of the ladder, but we
discuss also briefly the case of a uniform DM interaction
along the chains. Our motivation for restricting mainly to the case of
interactions along the rungs is the situation
in  ${\rm CaCu_2O_3}$ in which
the Cu--O--Cu bond angle in the ladder rungs is
123$^o$ indicating that in addition to the
Heisenberg coupling a Dzialoshinskii-Moriya interaction\cite{kiryukhin}
is also present in the rungs of the ladder. Note that a model of
coupled chains with Dzialoshinskii-Moriya interactions, motivated by
experiments on ${\rm CsCuCl_4}$ has also been considered in
Ref.~\onlinecite{bocquet_cscucl4}. However, in
Ref.~\onlinecite{bocquet_cscucl4}, the authors are considering an two
dimensional 
array of coupled chains using bosonization and RPA rather than an
isolated ladder.  
The paper is organized as
follows. We first outline the bosonization procedure of a weakly
coupled two leg spin ladder in presence of  DM and KSEA interactions
along the rungs. The derivation\cite{shelton_spin_ladders} of the low
energy theory is now quite standard, and the reader already familiar
with bosonization techniques should read it only to get acquainted
with our notations. Expressions for
the staggered correlation functions can then be obtained 
using  a mapping onto four
non-critical Ising models\cite{shelton_spin_ladders}. 
Then we study all uniform magnetic
correlation functions and calculate the temperature dependence of the
spin susceptibility and the NMR relaxation rate.  Finally we discuss
the very interesting effects of the anisotropic spin interactions on
the electron spin resonance (ESR) spectrum, where the DM interaction
triggers a phase with gapped but {\it incommensurate} correlation
functions with applied magnetic field perpendicular to the DM vector,
while the effect of a superimposed KSEA interaction is the
closure of the gap with a restoration of the $SU(2)$ symmetry.

\section{General considerations and bosonization}

\subsection{General considerations}
The Hamiltonian of an antiferromagnetic two leg ladder with both
Dzialoshinskii-Moriya (DM) and Kaplan-Shekhtman-Entin-Wohlman-Aharony (KSEA)
 interactions along the rungs is:

\begin{equation}
\label{ham} H=J_{\parallel}\sum_i ({\mathbf{S}}_{1,i}{\mathbf{S}}_{1,i+1}+
{\mathbf{S}}_{2,i}{\mathbf{S}}_{2,i+1})+J_{\perp} \sum_i
{\mathbf{S}}_{1,i}{\mathbf{S}}_{2,i}+{\mathbf{D}}\cdot \sum_i ({\mathbf{S}}_{1,i} \times
{\mathbf{S}}_{2,i}) + \sum_i  {\mathbf{S}}_{1,i}\tensor{A}{\mathbf{S}}_{2,i}
%\frac{1}{2J_{\perp}}\sum_i
%({\mathbf{S}}_{1,i}{\mathbf{D}})({\mathbf{S}}_{2,i}{\mathbf{D}}),
\end{equation}

\noindent where ${\mathbf{D}}$ is the  uniform Dzialoshinskii vector associated
with the bond between the spins on different chains. The last term in
the Hamiltonian (\ref{ham}) describes the
KSEA interaction, where $\tensor{A}$ is the anisotropy tensor.
Generally, in a system having full SU(2) symmetry, ${\mathbf{D}}$ and $\tensor{A}$ are not independent;
they are related to each other in such a way that a rotation of the spin axes 
maps the single-bond Hamiltonian onto an isotropic one with 
no preferred direction of
the staggered magnetization\cite{shekhtman_anisotropic_l}.
If we use the quantization axis ${\hat z}$ for the spins such that
${\mathbf{D}}=D{\hat z}$, the anisotropy tensor reduces to $A_{zz}=A$
and all other components become zero.
The Hamiltonian (\ref{ham}) reads:
\begin{eqnarray}
H= \sum_{i} J_\parallel ({\mathbf{S}}_{1,i}{\mathbf{S}}_{1,i+1}+
{\mathbf{S}}_{2,i}{\mathbf{S}}_{2,i+1}) + \frac{J_\perp+iD}{2}
S_{i,1}^+ S_{i,2}^- + \frac{J_\perp-iD}{2}S_{i,1}^- S_{i,2}^+ +
(J_\perp+A) S_{i,1}^z S_{i,2}^z
\end{eqnarray}
This Hamiltonian can be simplified by the gauge
transformation\cite{oshikawa_cu_benzoate}:
\begin{equation}\label{eq:gauge_transformation}
S_{i,1}^+=e^{-i\alpha} \tilde{S}_{i,1}^+ ; \mbox{ }\mbox{ } S_{i,2}^+=e^{i\alpha} \tilde{S}_{i,2}^+
\end{equation}
\noindent where $2\alpha=\arctan (D/J_\perp)$,
and brought to the form:
\begin{eqnarray}\label{tildeham}
H=\sum_{i} J_\parallel ({\mathbf{\tilde{S}}}_{1,i}{\mathbf{\tilde{S}}}_{1,i+1}+
{\mathbf{\tilde{S}}}_{2,i}{\mathbf{\tilde{S}}}_{2,i+1}) +\tilde{J}_\perp (\tilde{S}_{i,1}^x
\tilde{S}_{i,2}^x + \tilde{S}_{i,1}^y \tilde{S}_{i,2}^y) + (J_\perp+A) \tilde{S}_{i,1}^z \tilde{S}_{i,2}^z
\end{eqnarray}
\noindent where $\tilde{J}_\perp=\sqrt{J_\perp^2+D^2} {\rm
sign}(J_\perp)$.
The KSEA interaction is given by
$A=\sqrt{J_\perp^2+D^2}{\rm sign}(J_\perp)-J_\perp$ leading to the
restoration of ${\rm SU}(2)$ symmetry. For $D\ll J_\perp$, one
recovers the approximate expression $A\sim D^2/(2J_\perp)$ (see
Eq. (2.19) of Ref. \onlinecite{shekhtman_anisotropic_l}). 
Let us mention that  a Dzialoshinskii-Moriya interaction along the chain
direction:
\begin{equation}
H_{\text{DM}}^\parallel={\bf D}_\parallel \cdot \sum_{n,p} {\bf S}_{n,p} \times {\bf S}_{n+1,p}
\end{equation}
\noindent can also  be removed by another  gauge transformation:
\begin{eqnarray}\label{eq:gauge_parallel}
S^{+}_{n,p}=e^{i\delta n}\tilde{S}^{+}_{n,p},
\end{eqnarray}
where $\delta=\arctan(D_/parallel/J)$. 
This gauge transformation has no effect on the interchain coupling. However,
in the absence of KSEA interaction, it can induce an anisotropic
interaction along the chain direction. Further, this term induces
incommensuration in the correlation functions.
Let us discuss briefly the limit $J_\parallel=0$.
With Dzialoshinskii-Moriya interactions only, and for classical spins,
the spins
$\tilde{S}_{1,2}$ would lie in the ground state forming an angle of
$\pi$ with each other. Thus, the real spins would make an angle of
$\pi-2\alpha$ with each other. In that case, a weak ferromagnetic
moment proportional to $\sin \alpha$ would appear in the ground
state.
In the case of quantum spins $1/2$, we can rewrite the interchain
interaction up to an unimportant constant as:
\begin{equation}
H_{\text{interchain}}=\frac{\tilde{J}_\perp} 2
(\tilde{S_1}+\tilde{S_2})^2 +\frac{J_\perp -\tilde{J}_\perp} 2
(\tilde{S}_1^z+\tilde{S}_2^z)^2
\end{equation}
As a result, the total spin $\tilde{S}=\tilde{S}_1+\tilde{S}_2$ and the spin
projection along the z-axis $\tilde{S}^z=\tilde{S}_1^z+\tilde{S}_2^z$ are good
quantum numbers.
In the case of an antiferromagnetic interaction, the ground state is
the singlet state with $|\tilde{S}=0\rangle$ and the excited states
are $|\tilde{S}=1;\tilde{S}^z=\pm 1\rangle$ with energy
$E(\tilde{S}=1,\tilde{S}^z=\pm 1)=\frac{J_\perp+\tilde{J}_\perp}{2}$,
and $|\tilde{S}=1;\tilde{S}^z=0\rangle$  with energy
$E(\tilde{S}=1,\tilde{S}^z=0)=\tilde{J_\perp}$. We notice that the
state with $\tilde{S}^z=0$ is the highest excited state in this case.

For a ferromagnetic interaction, the state
$|\tilde{S}=1;\tilde{S}^z=0\rangle$ becomes the state of lowest
energy. The excited states are then $|\tilde{S}=1;\tilde{S}^z=\pm
1\rangle$ and $|\tilde{S}=0\rangle$, the latter one being the highest
excited state. Clearly, in the case of ferromagnetic interactions,
the system becomes equivalent to a XXZ spin 1 chain with easy plane
interaction $-K(S^z)^2$. This could lead to the observation of an XY2
phase\cite{schulz_spins}.

\subsection{Bosonization treatment}
For weak interchain coupling $J_\perp$ and ${\mathbf{D}}, \tensor{A}\ll
J_\parallel$, the Hamiltonian (\ref{tildeham}) can be bosonized following
Ref.\onlinecite{shelton_spin_ladders}. 

For the moment let us consider $\tensor{A}=0$. For $J_\perp=D=0$,  the low energy
properties of the Hamiltonian (\ref{tildeham}) are described in terms of
the boson operators $\phi_\alpha$ ($\alpha=1,2$) and their conjugate
momenta $\pi \Pi=\partial_x \theta_\alpha$. The Hamiltonian of chain
$\alpha$ is:
\begin{eqnarray}
H_\alpha=\int \frac{dx}{2\pi}u \left[ (\pi \Pi_\alpha)^2 + (\partial_x
\phi_\alpha)^2 \right]
\end{eqnarray}
where $u=\frac \pi 2 J_\parallel a $
is the ``spinon'' velocity and $a$ is the lattice spacing. Let us now
turn a weak non zero $J_\perp,D$ in the Hamiltonian (\ref{tildeham}).
By using the
continuum limit of spin operators\cite{shelton_spin_ladders}:
\begin{eqnarray}
\tilde{S}_\alpha^z(x=na)=\frac{\tilde{S}^z_{n,\alpha}}{a}=-\frac{\partial_x
\phi_\alpha}{\pi \sqrt{2}} +\frac{e^{i\frac{\pi x} a}}{\pi a}\lambda \sin
\sqrt{2} \phi_\alpha \nonumber \\
\tilde{S}_\alpha^+(x=na)=\frac{\tilde{S}^x_{n,\alpha}+i
\tilde{S}^y_{n,\alpha}}{a}=\frac{e^{i\sqrt{2}\theta_\alpha}}{\pi a} \left[
e^{i\frac{\pi x} a}\lambda + \cos \sqrt{2} \phi_\alpha \right],
\end{eqnarray}
\noindent we can rewrite (\ref{tildeham}) as:

\begin{eqnarray}
& & H=H_s+H_a \nonumber \\
& & H_s=\int \frac{dx}{2\pi} u\lbrack (\pi \Pi_s)^2+(\partial_x
\phi_s)^2\rbrack -\frac{J_{\perp}\lambda^2}{2\pi^2 a}\int dx \cos2\phi_s
\label{hs} \nonumber\\
& & H_a=\int \frac{dx}{2\pi} u\lbrack (\pi  \Pi_a)^2+(\partial_x
 \phi_a)^2\rbrack +\frac{J_{\perp}\lambda^2}{2\pi^2 a}\int dx
\cos2 \phi_a
+\frac{\tilde{J}_{\perp}\lambda^2}{\pi^2 a}\int dx \cos2 \theta_a \label{hatransf},
\end{eqnarray}

\noindent  where we have introduced the symmetric and antisymmetric
combinations:

\begin{equation}
\phi_s=\frac{\phi_1+\phi_2}{\sqrt{2}}, \mbox{ }\mbox{ }
\phi_a=\frac{\phi_1-\phi_2}{\sqrt{2}},
\end{equation}

\noindent and similar combinations for $\Pi$ and
$\theta$\cite{shelton_spin_ladders,strong_spinchains_long,orignac_2chain_long}.
$\lambda\simeq 1$ is the expectation value of the charge operator\cite{note}. Note that the
Hamiltonian $H_a,H_s$ can also be obtained by bosonizing directly the Hamiltonian
(\ref{ham}) and then performing a shift of the field $\theta_a$ which is equivalent to
the gauge transformation Eq. (\ref{eq:gauge_transformation}).
In the presence of a Dzialoshinskii-Moriya interaction along the
chains, the effects are two-fold. First, we have to make a shift
$\theta_p \to \theta_p +\delta \frac x a$, i.e. $\theta_s +
\sqrt{2}\beta \frac x a$. Since in the Hamiltonian~(\ref{hs}), only
the term $\cos 2\phi_s$ is present, such shift does not affect the
spectrum of the ladder. However, it induces incommensuration. The
spin-spin correlation function $\langle S^+(q)S^{-}(-q)\rangle$
 will present divergences for $q=\delta/a$
and $q=\pm \pi/a +\delta/a$ rather than respectively $q=0$ and $q=\pi/a$.
The second effect of the longitudinal Dzialoshinskii-Moriya
interaction is to induce an anisotropic interaction $\delta J^z
\sum_{i,p} S^z_{i,p} S^z_{i+1,p}$. Such interaction gives marginally
relevant interactions upon bosonization. On the other hand, the
interchain coupling gives relevant interactions of dimension 1. Thus,
for not too weak interchain coupling, the extra anisotropic
interaction will not affect the physical properties of the ladder, and
the only effect of the longitudinal Dzialoshinskii-Moriya interaction
will be a shift of the correlation functions.
 The Hamiltonians  $H_a,H_s$ can be further rewritten in terms of
non-interacting Majorana fermions\cite{shelton_spin_ladders}:

\begin{eqnarray}
\label{diracf}
\xi_{R,1}&=&\frac{\cos(\theta_s-\phi_s)}{\sqrt{\pi a}},\; 
\xi_{R,2}=\frac{\sin(\theta_s-\phi_s)}{\sqrt{\pi a}},\label{mais}  \\
\xi_{R,3}&=&\frac{\cos(\theta_a-\phi_a)}{\sqrt{\pi a}},\;
\xi_{R,4}=\frac{\sin(\theta_a-\phi_a)}{\sqrt{\pi a}}\label{maia},
\end{eqnarray}
and similar definitions for the $\xi_L$'s with $-\phi_{s,a} \to
\phi_{s,a}$. 
Reexpressed in terms of Majorana fermions the Hamiltonian reads:
\begin{eqnarray}
\label{hsmf} H = \sum_{a=1}^{4} \int dx \{ -\frac{iu}{2}(\xi^a_R \partial_x
\xi^a_R -\xi^a_L \partial_x \xi^a_L) -im_a \xi^a_R\xi^a_L \}.
\end{eqnarray}

\noindent The spectrum is thus composed of a Majorana doublet  ($\xi^1_{\nu}$, $\xi^2_{\nu}$),
($\nu=L,R$), with mass $m_{1,2}=m=\lambda^2 J_{\perp}/2\pi$  and two singlets $\xi^3_{\nu}$,
$\xi^4_{\nu}$ of respective masses
$m_3=\lambda^2(\frac{\tilde{J}_{\perp}}{\pi}-\frac{J_{\perp}}{2\pi})$,
$m_4=-\lambda^2(\frac{\tilde{J}_{\perp}}{\pi}+\frac{J_{\perp}}{2\pi})$. Without the DM
interaction\cite{shelton_spin_ladders} the spectrum would be formed of
a triplet $\xi^a_\nu$, $a=1,2,3$ with mass
$m$ and a singlet $\xi^4_\nu$ with a larger modulus mass
$3|m|$\cite{shelton_spin_ladders}. The loss of $SU(2)$ symmetry in the
presence of a DM interaction thus translates into the lifting of the
degeneracy of the triplet. Let us note that the Majorana
doublet describes the excitations of spin $\tilde{S}^z=\pm 1$, whereas the
singlet corresponds to the excitations of spin $\tilde{S}^z=0$. We
notice that the hierarchy of energy scales as well as the degeneracy
of excitations with $S^z=\pm 1$ that we had obtained in the strong
rung coupling limit are preserved in the limit of a small rung
coupling. In the presence of a KSEA interaction, the bosonization
treatment of (\ref{tildeham}) is identical to the one in
\onlinecite{shelton_spin_ladders}. We find that $m_{1,2,3}=\lambda^2 \tilde{J}_\perp/2\pi$, thus
recovering the triplet of the isotropic ladder.

\begin{figure}
\centerline{\epsfig{file=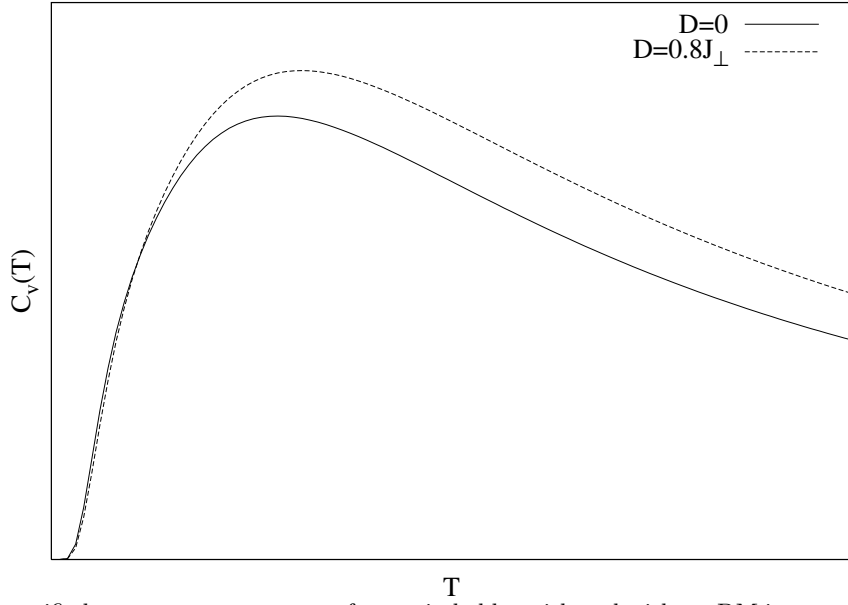,angle=-90, width=12cm}}
\caption{ A plot of specific heat versus temperature for a spin ladder
with and without DM interaction. A remarkable feature is
that at low temperature, the specific heat of the ladder with DM
interaction is lower while it is higher at high temperature.}
\label{fig:specheat}
\end{figure}

 From the Hamiltonian Eq. (\ref{hsmf}), the specific heat of the system
at low temperature is obtained in the form:
\begin{equation}
C_v\sim \left (\frac{m_1}{T}\right )^{3/2}
\frac{m_1}{\sqrt{2\pi}u}e^{-\frac{m_1}{T}}+ \left (\frac{m_2}{T}\right
)^{3/2} \frac{m_2}{\sqrt{2\pi}u}e^{-\frac{m_2}{T}}+\left (\frac{m_3}{T}\right
)^{3/2} \frac{m_3}{\sqrt{2\pi}u}e^{-\frac{m_3}{T}},
\end{equation}
\noindent If the KSEA interaction is different from zero, the SU(2) symmetry is restored and the
specific heat becomes, $C_v\sim \left
(\frac{\Delta''}{T}\right )^{3/2}
\frac{\Delta''}{\sqrt{2\pi}u}e^{-\frac{\Delta''}{T}}$,
%\chi\propto \sum_{i=1,2,3}\frac{\sqrt{2\pi m_i}}{u}T^{-1/2}e^{-m_i/T}.
%\end{equation}
where $\Delta''$ is the triplet mass $\lambda^2 \tilde{J}_\perp/2\pi$.
The difference between the regular ladder and the ladder with only DM
interaction is the contribution from the singlet
excitation with $S^z=0$. The specific heat as a function of
temperature is plotted on figure~\ref{fig:specheat}. A remarkable
feature is that at low temperature, the specific heat of the ladder
with DM interaction is lower while it is higher at high
temperature. 

\section{Staggered correlation functions}

To calculate the two-point correlation
functions of the
staggered magnetization, we use the mapping onto a
pair of non-critical 2d Ising
model introduced in
Refs.~\onlinecite{shelton_spin_ladders,gogolin_book}.
This mapping
permits to express the staggered correlation functions in terms of the
correlation functions
of the order and disorder parameters of the Ising model away from criticality
\cite{luther_ising,zuber_77,ogilvie_ising}.
To precise the correspondence, we will use the convention of
Ref.~\onlinecite{shelton_spin_ladders}  that
$J_\perp/t>0$, where $t$ is the reduced temperature of the
corresponding noncritical Ising model, $t=(T-T_c)/T_c$. With this convention,
 $J_\perp>0$ corresponds to the disordered
phase and  we have the following bosonization
formulas for order and disorder parameters
$\sigma_i$ and $\mu_i$ ($i=1,\ldots,4$):

\begin{eqnarray}
\cos(\phi_s)=\mu_1\mu_2, \mbox{ } \mbox{ }\sin(\phi_s)=\sigma_1\sigma_2, \nonumber \\
\cos(\theta_s)=\sigma_1\mu_2, \mbox{ } \mbox{
}\sin(\theta_s)=\mu_1\sigma_2,\nonumber \\
\cos (\phi_a)=\mu_3 \sigma_4, \mbox{ } \mbox{ }
\sin(\phi_s)=\sigma_3\mu_4,\nonumber \\
\cos(\theta_a)=\sigma_3\sigma_4 , \mbox{ } \mbox{ }
\sin(\theta_a)=\mu_3\mu_4.
\label{correspondence}
 \end{eqnarray}

Using formulas (\ref{correspondence}) along with the gauge
transformation (\ref{eq:gauge_transformation}), we obtain the
components of the total ${\mathbf{M}}$ and relative ${\mathbf{m}}$
staggered magnetization as:

\begin{eqnarray}
\label{magngreen}
& &M_x\sim (\cos \alpha \mbox{ }\sigma_1\mu_2\sigma_3\sigma_4 -\sin \alpha
\mbox{ }\sigma_1\mu_2\mu_3\mu_4)
\mbox{ } \mbox{ } m_x\sim
(\sin \alpha \mbox{ } \mu_1\sigma_2\sigma_3\sigma_4+\cos \alpha
\mbox{ }\mu_1\sigma_2\mu_3\mu_4 \nonumber) \\
& &M_y\sim (\cos \alpha \mbox{ }\mu_1\sigma_2\sigma_3\sigma_4 -\sin \alpha
\mbox{ }\mu_1\sigma_2\mu_3\mu_4)
\mbox{ } \mbox{ } m_y\sim
(\sin \alpha \mbox{ } \sigma_1\mu_2\sigma_3\sigma_4+\cos \alpha
\mbox{ }\sigma_1\mu_2\mu_3\mu_4) \nonumber \\
& &M_z\sim \sigma_1\sigma_2\mu_3\sigma_4 \mbox{ } \mbox{ } m_z\sim
\mu_1\mu_2\sigma_3\mu_4.
\end{eqnarray}
Comparing these expressions with those obtained by Shelton {\it et
al.}\cite{shelton_spin_ladders}, we see that the DM interaction mixes the total
and relative magnetization of the isotropic case in the $x-y$ plane,
while the $z$ components are unchanged. This is a consequence of the
easy-plane effect induced by the DM interaction. From
Eqs.(\ref{magngreen}) the two-point correlation functions
perpendicular and along the DM vector, are expressed in term of
correlation functions of the order and disorder
parameters\cite{shelton_spin_ladders,wu_ising}:
\begin{eqnarray}
& &\langle T_\tau M_x(x,\tau)
M_x(0,0)\rangle
=G_{\sigma}(\frac {mr} u )G_{\mu}(\frac{mr} u)
\lbrack \cos^2 \alpha G_{\sigma}(\frac{m_3 r} u)G_{\sigma}(\frac{m_4 r} u)+
\sin^2 \alpha G_{\mu}(\frac{m_3 r} u)G_{\mu}(\frac{m_4 r} u)\rbrack \nonumber \\
& &\langle T_\tau m_x(x,\tau)
m_x(0,0) \rangle
=G_{\sigma}(\frac{mr} u)G_{\mu}(\frac{mr} u)
\lbrack \sin^2 \alpha G_{\sigma}(\frac{m_3 r} u)G_{\sigma}(\frac{m_4 r} u)+
\cos^2 \alpha G_{\mu}(\frac{m_3 r} u)G_{\mu}(\frac{m_4 r} u)\rbrack \nonumber \\
& & \langle T_\tau M_z(x,\tau)
M_z(0,0) \rangle
=\lbrack G_{\sigma}(\frac{mr} u)\rbrack^2 G_{\mu}(\frac{m_3 r}
u)G_{\sigma}(\frac{m_4 r} u)\nonumber \\
& &\langle T_\tau m_z(x,\tau)
m_z(0,0) \rangle
=\lbrack G_{\mu}(\frac{mr} u)\rbrack^2 G_{\sigma}(\frac{m_3 r} u)G_{\mu}(\frac{m_4 r} u)\label{corrfunk}.
\end{eqnarray}
\noindent where $r=\sqrt{x^2+(u\tau)^2}$ and
$G_{\sigma}(m_ir/u)$ and $G_{\mu}(m_ir/u)$ are the
two-point correlation functions given by Eqs. (38) and (39) of
Ref.~\onlinecite{shelton_spin_ladders}.
  In order to
determine explicitly the correlations length in the direction
orthogonal and along the DM vector, we use the asymptotic expansions
of  the
functions\cite{abramowitz_math_functions}  $K_0(v)$
and $K_1(v)$ for $v \to \infty$:
\begin{eqnarray}
K_0(v)= \sqrt{\pi/(2v)}e^{-v}(1-1/(8v)+o(1/v)) \nonumber \\
\label{expan}
K_1(v)=\sqrt{\pi/(2v)}e^{-v}(1+3/(8v)+o(1/v)\ldots).
\end{eqnarray}

Substituting these expressions in Eq.(\ref{corrfunk}) we get, after
some cancellations:

\begin{eqnarray}
\langle T_\tau M_x(x,\tau)M_x(0,0)\rangle &\sim& \frac
{A_\infty^4 \sin^2 \alpha}{\sqrt{2\pi mr/u}} e^{-mr/u} \\
\langle T_\tau m_x(x,\tau)m_x(0,0)\rangle&\sim& \frac
{A_\infty^4 \cos^2 \alpha}{\sqrt{2\pi mr/u}} e^{-mr/u} \\
\langle T_\tau M_z(x,\tau)M_z(0,0)\rangle &\sim&
{A_\infty^4}\left(\frac{u}{2\pi (m^2 m_4)^{1/3} r}\right)^{3/2}
e^{-(2m+m_4) r/u} \\
\langle T_\tau m_z(x,\tau)m_z(0,0)\rangle &\sim& \frac
{A_\infty^4}{\sqrt{2\pi m_3 r/u}} e^{-m r/u}\\
\end{eqnarray}
\noindent We see that with only Dzialoshinskii-Moriya interactions,
the in-plane and out of plane correlation
functions have a different correlation length in contrast to the case
of the regular  ladder.  This
difference of correlation lengths should be observable in neutron
scattering experiments. The ratio of the correlation lengths should
give access to the intensity of the Dzialoshinskii-Moriya
interactions, since it is equal to $m_3/m=2\sqrt{1+(D/J_\perp)^2}-1$.
Another difference with the regular ladder is
that $\langle M_x M_x \rangle$ decays as slowly as $\langle m_x m_x
\rangle$ with DM interactions whereas its decay is much faster without
them. The ratio $\langle m_x m_x
\rangle/\langle M_x M_x \rangle$ provides a second measurement of
$D/J_\perp$.  When we take into account the KSEA
interaction, the $SU(2)$ symmetry is restored and correlation
functions have the same decay with $r$ in all directions with
$\xi=(\frac{\lambda^2 \tilde{J}_\perp}{2\pi u})^{-1}$. Nevertheless, in presence of
a KSEA interaction the decay of $\langle M_x M_x\rangle$ remains not faster than the
decay of $\langle m_x m_x\rangle$, and the ratio of these correlation functions gives
access to the strength of the Dzialoshinskii-Moriya interaction.
Using the above expressions~(\ref{corrfunk})
 we can easily calculate the imaginary
part of the dynamical transverse spin susceptibility for very small
momentum. We find:
\begin{eqnarray}
& & \Im
\chi^{xx}(q_\parallel,q_\perp=0,\omega)=\frac{A_\infty^4 \sin^2 \alpha
\pi^2u}{\pi\sqrt{(uq)^2+m^2}}
\left[ \delta(\omega-\sqrt{(uq)^2+m^2})
-\delta(\omega+\sqrt{(uq)^2+m^2})\right] \\
& & \Im
\chi^{xx}(q_\parallel,q_\perp=\pi,\omega)=\frac{A_\infty^4 \cos^2 \alpha
\pi^2u}{\pi\sqrt{(uq)^2+m^2}}
\left[ \delta(\omega-\sqrt{(uq)^2+m^2})
-\delta(\omega+\sqrt{(uq)^2+m^2})\right] \\
& & \Im
\chi^{zz}(q_\parallel,q_\perp=\pi,\omega)=\frac{A_\infty^4 \pi^2u}{\pi\sqrt{(uq)^2+m_3^2}}
\left[ \delta(\omega-\sqrt{(uq)^2+m_3^2})
-\delta(\omega+\sqrt{(uq)^2+m_3^2})\right]. \\
\end{eqnarray}

In presence of a KSEA interaction the
asymmetry disappears and the same optical magnon peak is obtained in
both directions at $\omega=\sqrt{(uq)^2+(\lambda^2 \tilde{J}_\perp/2\pi)^2}$. Thus
the analysis reveals that there is net way of determining the DM
interaction and the interplay between DM and KSEA interactions into
experiments.

\section{Magnetic susceptibility and NMR relaxation rate}\label{sec:chi_rmn}

\subsection{General formalism}\label{sec:formalism_green}
To  determine  the  temperature  dependence of the magnetic
susceptibility and of the NMR relaxation rate $1/T_1$, we only need
the slowly varying part of the spin density\cite{kishine_nmr}.
That component of the spin  density can be expressed in
terms of Majorana fermions. The corresponding correlators can the be
obtained by the approach of Ref.~\onlinecite{kishine_nmr}.
 The following  relations hold  for the
sum and  the difference of the spin  density currents of chains 1  and 2 along
the      spatial     directions     (1,2,3)(see      also     Appendix
\ref{appendix:currents}):
\begin{eqnarray}
(J_{\nu 1}+J_{\nu 2})^1=-i(\cos \alpha \mbox{ }\xi_\nu^2 \xi_\nu^3 - \sin \alpha\mbox{ } \xi_\nu^2
\xi^4_\nu) & &\mbox{ }\mbox{ }(J_{\nu 1} - J_{\nu 2})^1=-i(\cos \alpha \mbox{ }\xi_\nu^1
\xi^4_\nu + \sin \alpha \mbox{ }\xi_\nu^1 \xi_\nu^3 )\nonumber \\
(J_{\nu 1}+J_{\nu
2})^2=-i(\cos \alpha \mbox{ }\xi_\nu^3 \xi_\nu^1 + \sin \alpha \mbox{ }\xi_\nu^1 \xi^4_\nu) & & \mbox{
}\mbox{ } (J_{\nu 1} - J_{\nu 2})^2= -i(\sin \alpha \mbox{ }\xi_\nu^2 \xi_\nu^3 + \cos \alpha
\mbox{ }\xi_\nu^2 \xi^4_\nu )\nonumber
\\ \label{majorana}
(J_{\nu 1}+J_{\nu 2})^3=-i\xi_\nu^1 \xi_\nu^2 & &\mbox{ }\mbox{ } (J_{\nu 1} -
J_{\nu 2})^3=i \xi_\nu^3 \xi^4_\nu,
\end{eqnarray}
\noindent where $\nu=L,R$.
 From the previous expressions we evaluate the slowly varying component
of the Matsubara spin-spin correlation functions which are useful for
neutron scattering experiments and NMR relaxation rate:
\begin{equation}
\label{gensusc} \chi^{ab}(q,i\omega_n)=\int_0^\beta d\tau dx e^{i(\omega_n \tau -qx)}
\langle T_\tau J^a(x,\tau) J^b(0,0)\rangle_T \mbox{ }\mbox{ }\mbox{ }(a,b=1,2,3).
\end{equation}
\noindent The finite-temperature correlations are obtained by the
analytic continuation $i\omega_n\rightarrow \omega+i0_+$.  Note that
expression (\ref{gensusc}) corresponds to a tensor susceptibility, and
$J^{a(b)}$ is the total spin current.
\begin{equation}
\label{totj}
J^{a(b)}=\sum_{\nu=R,L}(J^{a(b)}_{\nu 1}+J^{a(b)}_{\nu 2}).
\end{equation}

The explicit expression for the correlation functions of the uniform currents in real
space is:
\begin{eqnarray}
\langle T_\tau J^1(x,\tau) J^1(0,0)\rangle& =& -\sum_{\nu=L,R \atop \mu=L,R} \lbrack
\cos^2 \alpha \langle T_\tau \xi^2_\nu \xi^3_\nu(x,\tau) \xi^2_\mu
\xi^3_\mu(0,0)\rangle +
\sin^2 \alpha \langle T_\tau \xi^2_\nu \xi^4_\nu(x,\tau) \xi^2_\mu \xi^4_\mu(0,0)\rangle \rbrack \nonumber \\
\langle T_\tau J^2(x,\tau) J^2(0,0)\rangle & =&\langle T_\tau
J^1(x,\tau) J^1(0,0)\rangle \nonumber \\
\label{corrreals}
\langle T_\tau J^3(x,\tau) J^3(0,0)\rangle & =& -\sum_{\nu=L,R \atop \mu=L,R}\langle
T_\tau \xi^1_\nu \xi^2_\nu(x,\tau) \xi^1_\mu \xi^2_\mu(0,0)\rangle,
\end{eqnarray}
\noindent the other contributions being zero.

From (\ref{corrreals}) we can evaluate the uniform
susceptibility by applying Wick's theorem.
The corresponding expressions can be written in compact form
introducing the thermal Green's
functions\cite{kishine_nmr}
for left- and right-moving triplet and singlet Majorana fermions:
\begin{eqnarray}
G^{\alpha}_{RR}(k,i\omega_n)& = & G^{\alpha}_{LL}(-k,i\omega_n)=-\frac{i\omega_n+uk}
{\omega_n^2+u^2k^2+m_\alpha^2} \nonumber \\
G^{\alpha}_{RL}(k,i\omega_n)& = & G^{\alpha\star}_{LR}(k,i\omega_n)=-\frac{im_\alpha}
{\omega_n^2+u^2k^2+m_\alpha^2},
\end{eqnarray}

\noindent where $\alpha$ stands for (1,2) in the case of the doublet
excitation or for (3,4) in case of a singlet excitation, and the
"polarization" function:
\begin{eqnarray}
\label{contr}
\Gamma^{\alpha \beta}(q,i\omega_n)=-\beta^{-1}\sum_{\omega_n'} \int \frac{dk}{2\pi}
\lbrack G_{RR}^\alpha(k,i\omega_n')G_{RR}^\beta(q-k,i\omega_n-i\omega_n')
+ G_{RL}^\alpha(k,i\omega_n')G_{RL}^\beta(q-k,i\omega_n-i\omega_n')\nonumber \\
+  G_{LR}^\alpha(k,i\omega_n')G_{LR}^\beta(q-k,i\omega_n-i\omega_n')+
  G_{LL}^\alpha(k,i\omega_n')G_{LL}^\beta(q-k,i\omega_n-i\omega_n')
\rbrack.
\end{eqnarray}

We find:
\begin{eqnarray}
& &\chi^{11}(q,i\omega_n)=  \left[\cos^2 \alpha \Gamma^{23}(q,\omega_n) -\sin^2
\alpha \Gamma^{24}(q,\omega_n)\right], \nonumber \\
& &\chi^{22}(i\omega_n)=
\chi^{11}(i\omega_n),\nonumber \\
& &\chi^{33}(i\omega_n)=  \Gamma^{12}(q,\omega_n) \\
& &\chi^{12}(i\omega_n)=
\chi^{21}(i\omega_n)=0.
\label{compchi}
\end{eqnarray}

\subsection{Calculation of the magnetic
susceptibility}\label{sec:susceptibility}
To obtain the static susceptibility, we need to take the limit $\omega
\to 0$, and the $k\to 0$ in $\chi(k,\omega)$.
 The evaluation of the Matsubara frequency sum for each term
of type (\ref{contr}) , in the limit $q\rightarrow
0$, leads to the following expression in the static limit. If
$m_\alpha = m_\beta$:
\begin{equation}
\label{gamma2}
\Gamma^{\alpha \alpha}(\omega \rightarrow 0)=\frac{1}{T} \int_0^\infty
\frac{dk}{2\pi}
\sech^2(\frac{\beta  \epsilon_{\alpha k}}{2}),
\end{equation}
\noindent where $\epsilon_{\alpha (\beta)k}=\sqrt{u^2k^2+m_{\alpha
(\beta)}^2}$.
For $m_\alpha\ne m_\beta$, a more complicated expression is obtained
(see Appendix~\ref{appendix:integral}). In this expression, a new
contribution can be isolated:
\begin{equation}
\label{gamma1}
\Gamma_{\alpha \beta}^0(q\to 0,\omega = 0)=\int \frac{dk}{2\pi}
\left(1-\frac{(uk)^2+m_\alpha
m_\beta}{\epsilon_\alpha(k)\epsilon_\beta(k)}\right) \frac{1-n_F(\epsilon_\alpha(k))-n_F(\epsilon_\beta(k))}{\epsilon_\alpha(k)+\epsilon_\beta(k)}.
\end{equation}
This contribution does not vanish for $T\to 0$, but gives:
\begin{equation}
\Gamma_{\alpha \beta}^0(q\to 0,\omega = 0)_{T=0}=\frac 1 {\pi u}
\left[ \frac 1 2 + \frac{m_\alpha m_\beta}{m_\alpha^2 -m_\beta^2} \ln
\left(\frac{m_\beta}{m_\alpha}\right)\right].
\end{equation}

\begin{figure}
\centerline{\epsfig{file=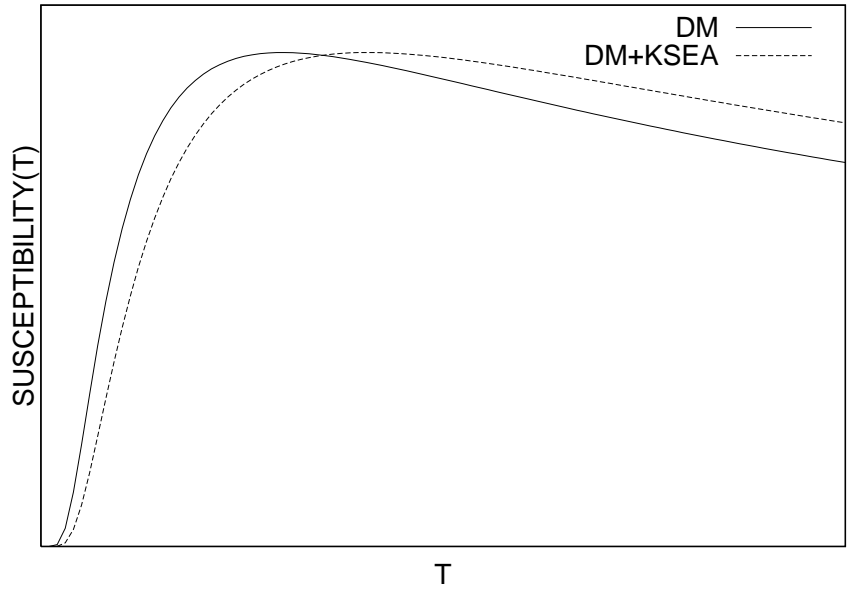,angle=-90, width=11.5cm}}
\caption{A plot of spin susceptibility along the $z$-axes versus
temperature for a spin ladder with DM interaction and DM plus KSEA
interaction, for $D/J_\perp=0.8$. The KSEA interaction restores the
SU(2) symmetry but a larger spectral gap appears in the low-T limit.}
\label{fig:susc}
\end{figure}

We obtain the final expressions of the
static and uniform components of spin susceptibility for $T\to 0$ as:

\begin{eqnarray}
& &\chi^{11}(0)=\chi^{22}(0)=\frac 1 {\pi u}\left\{ \cos^2 \alpha\left[ \frac 1 2 -
\frac{m m_3}{(m_3^2-m^2)}\ln(\frac{m_3}{m}) \right]
 +\sin^2 \alpha \left[ \frac 1 2 - \frac{m
m_4}{(m_4^2-m^2)}\ln(\frac{m_4}{m}) \right] + O(T^{-1/2} e^{-{\rm
min}\{m_i\}/T}) \right\}\nonumber \\
& & \chi^{33}(0)\simeq \frac{\sqrt{2\pi m}}{u}T^{-1/2}e^{-\frac{m}{T}}.
\end{eqnarray}
The result shows that the in-plane static susceptibility does not
vanish for $T\to 0$, whereas it decays exponentially,
$T^{-1/2}e^{-\lambda^2 J_\perp/(2\pi T)}$
along the DM vector. A similar result
holds in the case of the anisotropic
spin 1 chain where the
in-plane magnetic susceptibility is also finite at zero magnetic
field\cite{tsvelik_field}.Restoration
of the $SU(2)$ symmetry by the KSEA interaction
gives exponentially decaying susceptibilities in all
directions: 
$\chi^{11}(0)=\chi^{22}(0)=\chi^{33}(0)\simeq
T^{-1/2}e^{-\frac{\Delta''}{T}}$, where
$\Delta''=\lambda^2 \tilde{J}_\perp/2\pi$, similarly to an XXX ladder. This is
shown in Fig.\ref{fig:susc} where a remarkable feature is that at low
T the spin susceptibility is lower in presence of a DM plus KSEA
interaction, with a larger spectral gap opening.

\subsection{NMR relaxation rate}\label{sec:nmr_rate_majo}

In this section, we calculate the NMR longitudinal relaxation rate
$T_1^{-1}$. By  definition, $T_1^{-1}$ is related to
the imaginary part of the dynamical susceptibility
in the longitudinal direction:

\begin{equation}
(T_1^{-1}) \propto T \lim_{\omega \rightarrow 0} \sum_k \frac{\Im \chi(k,\omega)}{\omega},
\end{equation}
The susceptibility $\chi(k,\omega)$ can be decomposed into two
contributions, one form the uniform part of the magnetization,
$\chi_0(k,\omega)$ and one from the staggered component
$\chi_\pi(k,\omega)$.
 It is well
known that in the case of a ladder, the staggered component $\chi_\pi$
of the
spin density makes at low temperatures a negligible contribution to $T_1^{-1}$
compared with the uniform part\cite{ivanov_nmr_ladder_stag} $\chi_0$, so we can
focus on this latter contribution which is easily obtained from the
method of  Ref.~\onlinecite{kishine_nmr}. In the following, we will
denote $\chi_0(k,\omega)$ simply by $\chi$ to simplify notations.
We have:

\begin{equation}
\chi^a(k,i\omega_n)=\int_0^\beta e^{i \omega_n \tau} \langle\vec{J}_a(k,\tau)\vec{J}_a(0,0)\rangle,
\end{equation}

\noindent and $\vec{J}_a$ $(a=1,2)$ is the total spin current in chain 1 or
2. It can be derived by
Eq.s(\ref{symmetric_majorana})-(\ref{asymajor})-(\ref{asymajol}) in the Appendix
\ref{appendix:currents}. Using
the polarization function, we get the following expression for the
longitudinal susceptibility:

\begin{eqnarray}
\chi^a(q,i\omega_n)=& &[-\cos^2\alpha (\Gamma^{23}(q,i\omega_n)+\Gamma^{1 4}(q,i\omega_n)+
\Gamma^{31}(q,i\omega_n)+
\Gamma^{24}(q,i\omega_n))\nonumber \\
& &-\sin^2 \alpha (\Gamma^{24}(q,i\omega_n)+\Gamma^{13}(q,i\omega_n)+\Gamma^{14}(q,i\omega_n)+
\Gamma^{23}(q,i\omega_n))
\nonumber \\
& &-(\Gamma^{12}(q,i\omega_n)+\Gamma^{34}(q,i\omega_n))].
\end{eqnarray}

\noindent Taking into account that the fermionic branches $1,2$ have
the same spectrum, we can reduce the above expression to a simpler form:
\begin{eqnarray}
\chi^a(q,i\omega_n)=& &-[2(\Gamma^{23}(q,i\omega_n)+\Gamma^{14}(q,i\omega_n))+
(\Gamma^{12}(q,i\omega_n)+\Gamma^{34}(q,i\omega_n))].
\end{eqnarray}
The detailed  calculations are  reported in the
Appendix~\ref{appendix:integral}. Here, we only quote the final
results.
For $m_\alpha \ne m_\beta$, we have\cite{kishine_nmr}:
\begin{equation}
\label{gamdin1}
T\sum_q\lim_{\omega \rightarrow 0} \frac{\Im \Gamma^{\alpha \beta}(q,\omega)}{\omega} = \int_{{\rm max}(m_\alpha,m_\beta)}^\infty \frac{d\epsilon}{4\pi}
\left[ \frac{\epsilon^2+m_\alpha m_\beta}
{\sqrt{\epsilon^2-m_\alpha^2}\sqrt{\epsilon^2-m_\beta^2}}
\right] 
\sech^2(\frac{\beta  \epsilon}{2}).
\end{equation}

For $m_\alpha=m_\beta$, the NMR relaxation rate is
divergent\cite{sagi_nmr_haldane_gap} for $\omega=0$. For $\omega\sim
0$, the following expression is obtained:

\begin{equation}
\label{gamdin2}
T\sum_q \frac{\Im \Gamma^{\alpha \alpha}(q,\omega)}{\omega}\sim
\frac{m_\alpha} \pi e^{-m_\alpha/T} \ln \left(\frac{4e^\gamma
T}{\omega}\right),
\end{equation}
where $\gamma$ is Euler's constant\cite{abramowitz_math_functions}.

 From (\ref{gamdin1}) and (\ref{gamdin2}), the final expression for the
longitudinal relaxation rate reads:
\begin{eqnarray}
\frac 1 {T_1(\omega)} &=& \int_{m_3}^\infty \frac{d\epsilon}{4\pi}
\frac{\epsilon^2+mm_3}{\sqrt{\epsilon^2-m^2}\sqrt{\epsilon^2-m_3^2}}
\sech^2 \left(\frac{\beta\epsilon}2\right)+\int_{m_4}^\infty \frac{d\epsilon}{4\pi}
\left[\frac{\epsilon^2+mm_4}{\sqrt{\epsilon^2-m^2}\sqrt{\epsilon^2-m_4^2}}+\frac{\epsilon^2+m_3m_4}{\sqrt{\epsilon^2-m_3^2}\sqrt{\epsilon^2-m_4^2}}\right]
\sech^2 \left(\frac{\beta\epsilon}2\right) \nonumber \\
& + & \frac{m}\pi e^{-m/T} \ln
\left(\frac{4e^\gamma T}\omega\right)
\end{eqnarray}
\noindent The first two terms give the in-plane singlet-triplet
contributions, while the last two terms are the contributions along
the DM $\hat{z}$-axes. Compared with the results in
Ref.\onlinecite{kishine_nmr} the in-plane singlet-triplet
contributions are different.  For a DM plus KSEA interaction these
contributions become equal, leading to same result of
Ref.\onlinecite{kishine_nmr} with $J_\perp \rightarrow
\tilde{J}_\perp$, where an explicit dependence on DM interaction is
still present.

\section{Correlation functions in applied magnetic field and ESR spectra}

The electron-spin resonance (ESR), as well as neutron scattering
(NMR), have revealed powerful experimental techniques to test magnon
excitations in quasi-one-dimensional Heisenberg antiferromagnets. In
applied magnetic field, ESR experiments exhibit thermally activated
resonances, depending on the orientation, that could be interpreted as
transitions between magnon states whose energies are split by
magnetic fields and crystal-field anisotropy. The ESR power absorption
is proportional to the imaginary part of the dynamical susceptibility
$I(\omega)\propto \omega Im \chi(\omega)$. We calculate the ESR
transition frequencies in two cases, with applied magnetic field $h$
along the DM vector and perpendicular to it. The first case is
straightforward, the field couples to the doublet $\xi^1_\nu\xi^2_\nu$
and ESR intensity is non-zero only when $\omega \hbar=h$, where
$\omega$ is the microwave frequency. From the point of view of ESR
spectra the resonance has zero width. The result holds both in
presence and not of DM interaction, so that a measure of ESR spectra
with applied field along $\mathbf{D}$ does not permit to disregard the
anisotropic from the isotropic interactions. The analysis of ESR
spectra for the field perpendicular to DM vector is more involved. We
will begin with the case in which 
Dzialoshinskii-Moriya interactions exist only along the rungs. The
field dependence of the dispersion relation is no longer determined by
symmetry considerations alone. The Hamiltonian is given by:
\begin{equation}\label{eq:magfield_x}
H= -\frac{iu}{2} \sum_{a=1}^4\int dx (\xi^a_R\partial_x \xi^a_R - \xi^a_L\partial_x
\xi^a_L)- i \sum_{a=1}^4 m_a\int dx\xi^a_R\xi^a_L  +ih
\sum_{\nu=L,R}\int dx \lbrack \cos \alpha \xi^2_\nu\xi^3_\nu- \sin \alpha
\xi^2_\nu\xi^4_\nu\rbrack.
\end{equation}
Compared to the case without DM interaction, the field couples to both
singlets $\xi^3$ and $\xi^4$ so that also singlet modes participate to
ESR resonances. To diagonalize the Hamiltonian
(\ref{eq:magfield_x}) we express it in Fourier space as:

\begin{equation}
H=\sum_{k>0} {}^t\vec{\zeta}(-k) {\cal H}(k) \vec{\zeta}(k)
\end{equation}
\noindent where:
\begin{equation}
\vec{\zeta}(k)=\left(
\begin{array}{c}
\xi_R^2(k) \\
\xi_L^2(k) \\
\xi_R^3(k) \\
\xi_L^3(k) \\
\xi_R^4(k) \\
\xi_L^4(k) \end{array} \right)=\frac 1 {\sqrt{L}} \int_0^L dx
\vec{\zeta}(x) e^{-ikx},
\end{equation}
\noindent and:
\begin{equation}
{\cal H}(k)=\left(\begin{array}{cccccc}
uk & im & ih \cos \alpha & 0 & ih \sin \alpha & 0 \\
-im & -uk & 0 & ih \cos \alpha & 0 & ih \sin \alpha \\
-ih \cos \alpha & 0 & uk & im_3 & 0 & 0 \\
0 & -ih \cos \alpha & -im_3 & -uk & 0 & 0 \\
-ih \sin \alpha & 0 & 0 & 0 & uk & im_4 \\
 0 &- ih \sin \alpha&  0 & 0 &  -im_4 & -uk \end{array} \right).
\end{equation}

\begin{figure}
\centerline{\epsfig{file=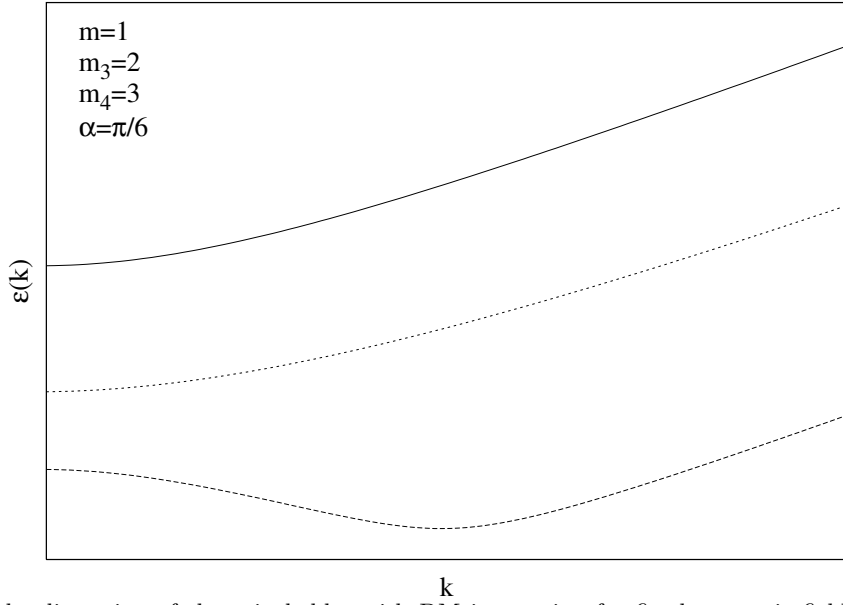,angle=-90,width=12cm}}
\caption{A plot of the dispersion of the spin ladder with DM
interaction for fixed magnetic field $h=3$, $u=1$; $m$ is the Majorana doublet mass,
$m_{3,4}$ are the  masses of the singlets . We notice that
there is no closure of any gap upon application of a quite strong
field. From the dispersion, we should expect gapped but incommensurate
correlations to develop in the system.}
\label{fig:algap}
\end{figure}

\begin{figure}
\centerline{\epsfig{file=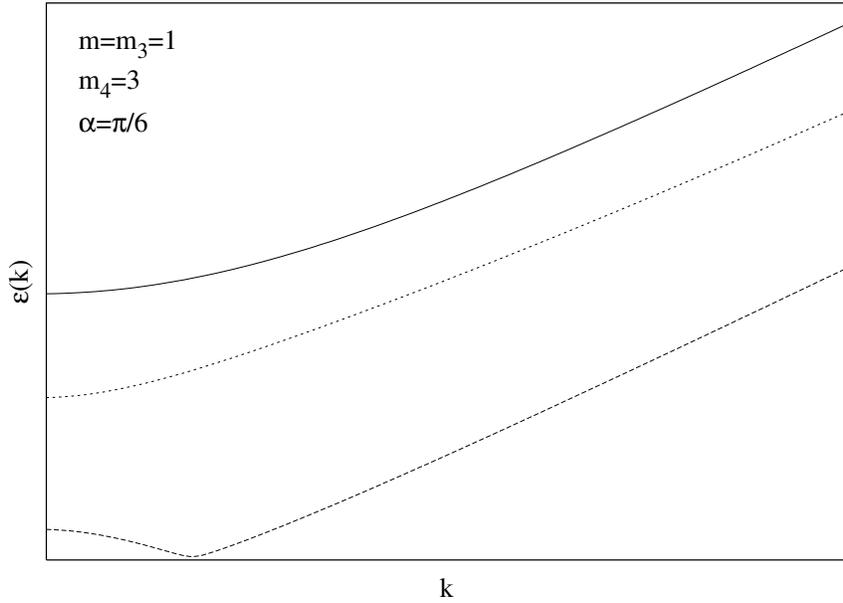,angle=-90,width=12cm}}
\caption{A plot of the dispersion of the spin ladder with DM+KSEA
interaction for fixed magnetic field $h=1.5$, $u=1$. We notice
a closure of the gap upon application of the magnetic
field. Thus, the KSEA interaction makes the $\hat{x}$ axis equivalent
to the $\hat{z}$ axis. }
\label{fig:nogap}
\end{figure}

\noindent The spectrum of ${\cal H}$ is obtained by solving the equation
${\rm det}(\epsilon -{\cal H})=0$.
This equation reduces to the form:

\begin{equation}
\label{pol3ord}
y^3+a_2y^2+a_1y+a_0=0
\end{equation}

\noindent where $y=\epsilon(k)^2$, and whose coefficients are listed in the
Appendix \ref{appendix:coeff}. In presence of the KSEA interaction we have a
slight simplification coming from the condition $m_1=m_2$. This permits
a reduction of the matrix to blocks simplifying the
diagonalization.  Some asymptotic expressions for the spectrum are obtained in the large
and small $h$ limit and they are explicitly described in the Appendix
\ref{appendix:coeff}. As discussed in Ref.\onlinecite{affleck_esr},
the ESR signal will result from transitions between the two lowest
branches $\epsilon_i(k,\alpha,h)$ $(i=1,2)$.
Transitions will occur at values of field and wavevector
satisfying $|\epsilon_1(k,\alpha,h)-\epsilon_2(k,\alpha,h)|=\omega
\hbar$, for a fixed DM interaction. In the anisotropic case the
difference will depend on momentum $k$, and $\alpha$ causing a resonance
broadening. Comparing with the results of ESR analysis of the
one-dimensional Heisenberg antiferromagnet\cite{affleck_esr}, in
presence of a pure DM interaction the three magnon branches mix even
for a field along one of the symmetry axes due to the easy-plane
effect. If DM interaction would not be present a branch would be
unaffected by the external field, and only the other two would
split. ESR experiments with field orientation along symmetry axes
would help to confirm the picture of a DM induced mixing. As shown in
Fig.~\ref{fig:algap}, under the application of the magnetic field
perpendicular to the DM vector, a remarkable result is that in
contrast to the case of an isotropic ladder, the magnetic field does
not close the gap, thus gapped but {\it incommensurate} correlations
develop.  This result is in consistent agreement with neutron
diffraction studies \cite{kiryukhin} on ${\rm CaCu_2O_3}$, whose structure
is similar to that of prototype two-leg spin ladder compound
${\rm SrCu_2O_3}$.  The results show that the magnetic structure is
incommensurate in the direction of the frustrated interchain
interaction in presence of an applied magnetic field along the $a$-axes.  An
evaluation of the DM interaction reveals that DM coupling in this
system may be as large as several meV.  In presence of a KSEA
interaction, the numerical results show a closure of the gap upon
application of the magnetic field. The KSEA interaction makes the axis
parallel to $\mathbf{D}$ equivalent to the one perpendicular to it,
restoring the $SU(2)$ symmetry. This result is shown in
Fig.~\ref{fig:nogap}.  The most general orientation of the magnetic
field in the plane perpendicular to $\mathbf{D}$ will be dealt
elsewhere. 

We now briefly discuss  the case in which we have 
Dzialoshinskii-Moriya interaction both along the rung and along the
chain direction. Then, we have to perform a gauge transformation
(\ref{eq:gauge_parallel}). As a result, the coupling to the field
parallel to $x$ direction becomes:
\begin{equation}
H_{\text{field}}=-h \sum_{n} (\cos(\delta n) S_n^x -\sin(\delta n) S_n^y).
\end{equation}
Using the expressions (\ref{majorana}) of the uniform component of
the spin density, we obtain:
\begin{equation}
H_{\text{field}}=-i h \sum_\nu \int dx (\cos \alpha \xi_\nu^3 -\sin
\alpha \xi_\nu^4) \left[ \cos \left(\frac{\delta x}{a}\right)
\xi_\nu^2 + \sin \left(\frac{\delta x}{a}\right) \xi_R^1\right].
\end{equation}
 Thus, the magnetic field is now acting
as a periodic potential on the Majorana fermions and
band structure in the spectrum should be expected. 
The discussion of
the effects of such band structure on ESR response will be discussed
elsewhere\cite{preprint}.

\section{Conclusions}

We have presented a Majorana fermion description of spin 1/2 ladders in
presence of anisotropic superexchange spin interactions along the
rungs which allows for the calculation of static and dynamic magnetic
properties. We have shown that in presence of a pure DM interaction,
the systems always has a spectral gap and the lower lying excitations
are Majorana doublets, instead of triplets for the regular ladder, 
indicating that the DM
interaction alone is acting as an easy-plane effect.  We have computed
both the asymptotic behavior and temperature dependence of the
anisotropic spin susceptibility helpful for NMR experiments. Our
analysis reveals that the easy-plane effect induced by a pure DM
interaction gives shorter correlation lengths along the DM vector,
compared to the orthogonal direction. The symmetric counterpart of the
DM interaction, i.e. the KSEA interaction, restores the SU(2) symmetry
and the same correlation length is found in all direction. We have
demonstrated that the interplay of the DM interaction and KSEA
interaction results in very interesting behaviors of the static
susceptibility and ESR spectra that are of direct experimental
relevance. In presence of a pure DM interaction gapped but {\it
incommensurate} correlations develop while the effect of a
superimposed KSEA interaction is the closure of the gap when an external
magnetic field is applied along one of the symmetry axes. The overall
analysis reveals that there is a net way of determining the DM
interaction and the interplay of the DM and KSEA interactions in
experiments.  It would be interesting to extend the present analysis
to the strong-coupling limit, or analyse the effect of a canted
antiferromagnetism over the DM and KSEA interactions.

\section{Acknowledgments}

We thank N. Andrei and R. Chitra  for useful discussions and careful
reading of the paper.

\newpage

\appendix

\section{Uniform components of the correlation functions}\label{appendix:currents}

According to bosonization, we can decompose the spin density into a
uniform and a staggered component as:
\begin{eqnarray}
{\bf S}_p ={\bf J}_{R,p} + {\bf J}_{L,p} + e^{i\pi x/a}
{\bf n}_p
\end{eqnarray}
We have discussed the staggered component in the previous sections. In
the present section,  we derive the expression of the uniform
component of the spin density as a function of the Majorana fermion
operators.
Using the definition of the currents\cite{shelton_spin_ladders} and
the transformation (\ref{eq:gauge_transformation}), we obtain the
following expressions for the uniform component of the spin density:
\begin{eqnarray}
& &(J_{R1}+J_{R2})^1=-i(\cos \alpha \xi_R^2 \xi_R^3 - \sin \alpha \xi_R^2
\xi_R^4) \nonumber \\
& &(J_{R1}+J_{R2})^2=-i(\cos \alpha \xi_R^3 \xi_R^1 + \sin \alpha \xi_R^1
\xi_R^4) \nonumber \\
\label{symmetric_majorana}
& &(J_{R1}+J_{R2})^3=-i\xi_R^1 \xi_R^2 \nonumber \\
\end{eqnarray}
And:
\begin{eqnarray}
& &(J_{L1}+J_{L2})^1=-i(\cos \alpha \xi_L^2 \xi_L^3 -  \sin \alpha \xi_L^2 \xi_L^4)
\nonumber \\
& &(J_{L1}+J_{L2})^2=-i(\cos \alpha \xi_L^3 \xi_L^1 +  \sin \alpha \xi_L^1 \xi_L^4)
\nonumber \\
& &(J_{L1}+J_{L2})^3=-i\xi_L^1 \xi_L^2 \nonumber \\
\end{eqnarray}

For the differences of the currents, we obtain:
\begin{eqnarray}\label{asymajor}
& &(J_{R1} - J_{R2})^1=-i(\cos \alpha \xi_R^1 \xi_R^4 + \sin \alpha \xi_R^1 \xi_R^3 )\nonumber \\
& &(J_{R1} - J_{R2})^2= -i(\sin \alpha \xi_R^2 \xi_R^3 + \cos \alpha \xi_R^2 \xi_R^4 )\nonumber \\
& &(J_{R1} - J_{R2})^3=i \xi_R^3 \xi_R^4
\end{eqnarray}

\begin{eqnarray}\label{asymajol}
& &(J_{L1} - J_{L2})^1=-i(\cos \alpha \xi_L^1 \xi_L^4 + \sin \alpha \xi_L^1 \xi_L^3 )\nonumber \\
& &(J_{L1} - J_{L2})^2=-i(\sin \alpha \xi_L^2 \xi_L^3 + \cos \alpha \xi_L^2 \xi_L^4)  \nonumber \\
& &(J_{L1} - J_{L2})^3=i \xi_L^3 \xi_L^4.
\end{eqnarray}

 From the previous expressions we evaluate the slowly varying component
of the Matsubara spin-spin correlation functions which are useful for
neutron scattering experiments and NMR relaxation rate as reported in
the main text.

\section{Details on the calculation of the NMR relaxation rate and
magnetic susceptibility}\label{appendix:integral}

To obtain the NMR relaxation rate or the magnetic susceptibility we
have to calculate the
following quantity

\begin{equation}
\label{integral}
T\int \frac{dq}{2\pi} \lim_{\omega \rightarrow 0} \frac{\Im \Gamma^{\alpha \beta}(q,\omega)}{\omega},
\end{equation}

\noindent where   $\Gamma^{\alpha \beta}(q,\omega)$ is defined in
Eq.~(\ref{contr}).

We use the following decompositions:
\begin{eqnarray}\label{eq:dees_green_func}
G_{RR}^\alpha(k,i\omega_n)=-\frac 1 2
\left[\left(1+\frac{uk}{\epsilon_\alpha(k)}\right) \frac 1 {i\omega_n +
\epsilon_\alpha(k)} + [\left(1-\frac{uk}{\epsilon_\alpha(k)}\right) \frac 1 {i\omega_n -
\epsilon_\alpha(k)} \right] \\
G_{RL}^\alpha(k,i\omega_n)=\frac{-im}{2\epsilon_\alpha(k)}\left[\frac 1 {i\omega_n
+\epsilon_\alpha(k)} - \frac{1}{i\omega_n - \epsilon_\alpha(k)} \right]
\end{eqnarray}

To obtain after performing the Matsubara sum:
\begin{eqnarray}\label{eq:polarization_matsubara}
\Gamma_{\alpha \beta}(q,\omega_n)& = & \Gamma^{pp}_{\alpha
\beta}(q,\omega_n) + \Gamma^{ph}_{\alpha \beta}(q,\omega_n), \\
\Gamma^{pp}_{\alpha \beta}(q,\omega_n)&=&\int \frac{dk}{4\pi}
\left(1-\frac{u^2 k(k-q)+m_\alpha m_\beta}{\epsilon_\alpha(k)
\epsilon_\beta(k-q)}\right)
(1-n_F(\epsilon_\alpha(k))-n_F(\epsilon_\beta(k-q))) \nonumber \\
&\times & \left[\frac 1
{i\omega_n + \epsilon_\alpha(k) +\epsilon_\beta(k-q)}
 -  \frac 1
{i\omega_n - \epsilon_\alpha(k) -\epsilon_\beta(k-q)}\right],
\label{eq:gamma_pp} \\
\Gamma^{ph}_{\alpha \beta}(q,\omega_n)&=&\int \frac{dk}{4\pi} \left(1+\frac{u^2 k(k-q)+m_\alpha m_\beta}{\epsilon_\alpha(k)
\epsilon_\beta(k-q)}\right)
(n_F(\epsilon_\alpha(k))-n_F(\epsilon_\beta(k-q)))\nonumber \\ &\times &
\left[ \frac 1
{i\omega_n +\epsilon_\beta(k-q) - \epsilon_\alpha(k)} - \frac 1
{i\omega_n +\epsilon_\alpha(k) -
\epsilon_\beta(k-q)}\right]. \label{eq:gamma_ph}
\end{eqnarray}
\subsection{NMR rate}
From  Equation (\ref{eq:polarization_matsubara}), it is clear that only
$\Gamma^{ph}(q,\omega_n)$  can make a
non-zero contribution to $\Im \Gamma_{\alpha \beta}$ for $\omega \to 0$.
Making the analytic continuation $i\omega_n \to \omega+i0$ and
considering the imaginary part, we obtain:
\begin{eqnarray}\label{eq:imaginary_gamma}
\Im \Gamma^{ph}_{\alpha \beta}(q,\omega)& = &\int \frac{dk}{4} \left(1+\frac{u^2 k(k-q)+m_\alpha m_\beta}{\epsilon_\alpha(k)
\epsilon_\beta(k-q)}\right)
(n_F(\epsilon_\alpha(k))-n_F(\epsilon_\beta(k-q)))\nonumber \\ &\times
&\left[
\delta\left(\omega -\epsilon_\beta(k-q) + \epsilon_\alpha(k)\right) -
\delta\left(\omega -\epsilon_\alpha(k) +  \epsilon_\beta(k-q)\right) \right]
\end{eqnarray}

Performing the integration over $q$, and making the variable change
$k-q \to k'$, we obtain the following expression for the contribution
of $\Gamma^{ph}_{\alpha \beta}$ to the NMR rate:

\begin{eqnarray}
\frac T {8\pi} \int dk \int dk' \left(1 +\frac{u^2 k k' + m_\alpha
m_\beta}{\epsilon_\alpha(k)
\epsilon_\beta(k')}\right)(n_F(\epsilon_\alpha(k))-n_F(\epsilon_\beta(k')))
\frac{\delta(\omega -\epsilon_\beta(k') + \epsilon_\alpha(k)) -
\delta(\omega -\epsilon_\alpha(k) +  \epsilon_\beta(k'))}\omega
\end{eqnarray}
The term containing $k k'$ vanishes by symmetry. Since the rest is
invariant under $k \to -k$ or $k' \to -k'$, we can reduce the
integration on $(k,k')$ to $[0,\infty]\times[0,\infty]$ and multiply
by a factor of $4$.
It is now convenient to introduce the variable change:
$\epsilon=\epsilon_\alpha(k)$, $\epsilon'=\epsilon_\beta(k')$ to find
that the contribution to the NMR rate reads:
\begin{eqnarray}
\frac{T}{2\pi} \int_{m_\alpha}^\infty
\frac{d\epsilon}{\sqrt{\epsilon^2 -m_\alpha^2}}   \int_{m_\beta}^\infty
\frac{d\epsilon'}{\sqrt{(\epsilon')^2 -m_\beta^2}}
(\epsilon\epsilon'+m_\alpha m_\beta) (n_F(\epsilon)-n_F(\epsilon'))
\frac{\delta(\omega+\epsilon-\epsilon')-\delta(\omega+\epsilon'-\epsilon)}{\omega}
\end{eqnarray}

For $m_\alpha > m_\beta$ and $\omega>0$, this expression is
rewritten\cite{kishine_nmr}:
\begin{equation}
\lim_{\omega \to 0} \frac T \omega \Im \Gamma^{ph}_{\alpha \beta}(q,\omega)
=\frac{1}{4\pi} \int_{m_\alpha}^\infty d\epsilon\frac{\epsilon^2
+m_\alpha m_\beta}
{\sqrt{(\epsilon^2-m_\alpha^2)(\epsilon^2-m_\beta^2)}}
\cosh^{-2}(\frac{\epsilon}{2T})
\end{equation}

For $m_\alpha=m_\beta$, some extra care is needed as the NMR
relaxation rate contains a logarithmic divergence for $\omega \to 0$.
This time, the contribution to the NMR relaxation rate is:
\begin{equation}
\frac T \omega \Im \Gamma^{ph}(q,\omega)=\frac 1 {4\pi} \int_{m_\alpha}^\infty
d\epsilon \frac{\epsilon^2 + m_\alpha^2}{\sqrt{\epsilon^2 -
m_\alpha^2} \sqrt{(\epsilon+\omega)^2 -m_\alpha^2}} \cosh^{-2}
\left(\frac{\epsilon}{2T}\right)
\end{equation}

Since the divergence comes from $\epsilon\sim m$, to estimate the
singular part we can approximate
the integral by:
\begin{equation}
\frac{ m_\alpha} {\pi} \int_{m_\alpha}^\infty \frac{
e^{-\epsilon/T} d\epsilon}{\sqrt{(\epsilon -m_\alpha)
(\epsilon+\omega -m_\alpha)}}
\end{equation}
Provided $m\gg T$.
With the variable change:
\begin{equation}
\epsilon=m+\frac \omega 2 (\cosh t -1)
\end{equation}
The integral is rewritten:
\begin{equation}\label{eq:t1_equal_mass}
\frac{ m_\alpha} {\pi} e^{-(m_\alpha-\omega/2)/T} \int_0^\infty dt
e^{-\frac{\omega}{2T}\cosh t}\sim \frac {m_\alpha}{\pi} e^{-m/T}
\left[
K_0\left(\frac{\omega}{2T}\right) +o(1) \right]
\end{equation}

Expansion of the Bessel function for $\omega \to 0$ leads to the
formula (3.9) of Ref. \onlinecite{sagi_nmr_haldane_gap}.

\subsection{Susceptibility}

The calculation of the susceptibility for $m_\alpha=m_\beta$ is
straightforward. To obtain the susceptibility for $m_\alpha \ne
m_\beta$, we must note that there are two contributions, one from
$\Gamma^{pp}$, the other from $\Gamma^{ph}$. Interestingly, the
contribution for $\Gamma^{pp}$ does not vanish for $T=0$.
Below, we give a detailed calculation of $\Gamma^{pp}$. Starting from
Eq.~(\ref{eq:gamma_pp}) for $\omega_n=0$, we obtain:
\begin{eqnarray}
\Gamma^{pp}(q\to 0,\omega_n=0)=\int \frac{dk}{2\pi}
\left(1-\frac{(uk)^2+m_\alpha m_\beta}{\epsilon_\alpha(k)
\epsilon_\beta(k)}\right) \frac
{1-n_F(\epsilon_\alpha(k))-n_F(\epsilon_\beta(k)}{\epsilon_\alpha(k)+\epsilon(\beta(k))}
\end{eqnarray}

Multiplying numerator and denominator by
$\epsilon_\alpha-\epsilon_\beta$, and taking $T=0$, we obtain:
\begin{eqnarray}\label{eq:suscep_diff_mass}
\Gamma^{pp}(q\to 0,\omega_n=0)=\frac 1 {m_\alpha^2-m_\beta^2} \int
\frac{dk}{2\pi}\left[\epsilon_\alpha(k) + \frac{(uk)^2+m_\alpha
m_\beta}{\epsilon_\alpha(k)} - \epsilon_\beta(k) - \frac{(uk)^2+m_\alpha
m_\beta}{\epsilon_\beta(k)}\right].
\end{eqnarray}

To calculate the integral~(\ref{eq:suscep_diff_mass}), we consider:
\begin{equation}
\int_0^{K} \frac{dk}{2\pi} \left(\epsilon_\alpha(k) +
\frac{(uk)^2+m_\alpha m_\beta}{\epsilon_\alpha}\right),
\end{equation}
\noindent where $K$ is a cut-off.
Using the variable change: $uk=m_\alpha \cosh \theta$, we find:
\begin{equation}
\int_0^{K} \frac{dk}{2\pi} \left(\epsilon_\alpha(k) +
\frac{(uk)^2+m_\alpha m_\beta}{\epsilon_\alpha}\right)=\frac 1 {\pi u}
\left[ uK \sqrt{(uK)^2+m_\alpha^2}+ m_\alpha m_\beta \ln
\left(\frac{2uK}{m_\alpha}\right) \right].
\end{equation}
With this result, we find that:
\begin{equation}
\Gamma^{pp}(q\to 0,\omega_n=0)=\frac{1}{\pi u (m_\alpha^2 -m_\beta^2)}
\lim_{K\to \infty} \left[\frac{uK
(m_\alpha^2-m_\beta^2)}{\sqrt{(uK)^2+m_\alpha^2}+\sqrt{(uK)^2+m_\beta^2}}
+ m_\alpha m_\beta \ln \left(\frac{m_\beta}{m_\alpha}\right)\right],
\end{equation}
\noindent i.e.
\begin{equation}\label{eq:suscep_dm_final}
\Gamma(q=0,\omega=0)=\frac 1 {\pi u} \left[ \frac 1 2 + \frac{m_\alpha
m_\beta}{m_\alpha^2 - m_\beta^2} \ln \left(\frac{m_\beta}
{m_\alpha}\right)\right].
\end{equation}

\section{Energy levels of the DM ladder in a magnetic field}\label{appendix:coeff}

The secular equation that gives the eigenvalues of the Hamiltonian
(\ref{eq:magfield_x}) is of the form:

\begin{equation}
\label{pol3or}
y^3+a_2y^2+a_1y+a_0=0
\end{equation}

\noindent where $y=\epsilon(k)^2$.
We put $\epsilon_i(k)=k^2+m_i^2$. The coefficients $a_i$ are given by

\begin{eqnarray}
a_2 &=& -(\epsilon_1^2+\epsilon_2^2+\epsilon_3^2+2h^2) \nonumber \\
a_1&=& h^4+2h^2 A({\mathbf{m}},\alpha) +
\epsilon_1^2\epsilon_2^2+\epsilon_1^2\epsilon_3^2+\epsilon_2^2\epsilon_3^2
\nonumber \\
a_0&=& -h^4 B(k,{\mathbf{m'}},\alpha)+ 2h^2 C(k,{\mathbf{m}},\alpha)-\epsilon_1^2\epsilon_2^2\epsilon_3^2.
\end{eqnarray}

\begin{eqnarray}
\label{coeff} A({\mathbf{m}},\alpha)&=& (m_4^2\cos^2 \alpha
-m_1m_4\sin^2 \alpha +m_3^2\sin^2 \alpha -m_1m_3\cos^2 \alpha )\nonumber \\
B(k,{\mathbf{m'}},\alpha)&=& (m_4^2\cos^4 \alpha +2m_3 m_4\sin^2
\alpha \cos^2 \alpha  +m_3^2\sin^4 \alpha +k^2) \nonumber \\
C(k,{\mathbf{m}},\alpha)&=& (m_1m_3m_4^2 \cos^2 \alpha +k^2
m_4^2\cos^2 \alpha+m_1m_3^2m_4 \sin^2 \alpha \nonumber \\ &+& k^2 m_1m_4 \sin^2
\alpha+k^2 m_3^2 \sin^2 \alpha+k^2 m_1m_3 \cos^2 \alpha+k^4),
\end{eqnarray}
\noindent where ${\mathbf{m}}=(m_1,m_3,m_4)$ and ${\mathbf{m'}}=(m_3,m_4)$.
For some specific values of $h$ the solutions of (\ref{pol3or})
are plotted in Figs.\ref{fig:algap}-\ref{fig:nogap}. For small $h$ or large $h$ we derive
the following asymptotic expressions for the solutions:

\subsection{Behavior in strong magnetic field}

For strong magnetic field, it is convenient to search for solutions of
the form: $y(k)=h^2 u(k)$.
Doing such replacement, we obtain the equation for $u$ in the form:
\begin{equation}
u(u-1)^2 - \frac{\epsilon_1^2+\epsilon_2^2+\epsilon_3^2}{h^2} u^2 +
 \left( \frac {2A}{h^2}+ \frac{\epsilon_1^2\epsilon_2^2+\epsilon_2^2\epsilon_3^2+
 \epsilon_1^2\epsilon_3^2}{h^4}\right)u -\left(\frac B {h^2}+ \frac C {h^4} -
 \frac{\epsilon_1^2 \epsilon_2^2 \epsilon_3^2}{h^6}\right)=0
\end{equation}
Thus, one can search for solutions of the form:
\begin{eqnarray}
u= 1+E/h +O(1/h^2) \\
u=E/h^2+O(1/h^2)
\end{eqnarray}
In particular, we find the solution $u=B/h^2$, which leads to
$\epsilon(k)=\sqrt{B}$ which is an effectively non-magnetic mode. This
non-magnetic mode can be obtained in the numerical calculations for
high magnetic field ($h\ge 9$). We note that for $\alpha=0$ we recover
the singlet mode of the spin ladder. The other solution leads to an
energy dispersion of the form:
\begin{equation}
\epsilon(k)=\pm \gamma(k/2,{\mathbf{m}},\alpha) \pm h,
\end{equation}

\noindent where  $\gamma(k/2,{\mathbf{m}})=\sqrt{k^2+(m+m_3 \cos^2 \alpha
+m_4 \sin^2 \alpha)^2/4}$.

We note that for $\alpha=0$ and $m=m_3$ we recover the magnetic
modes of the spin ladder.  So, in high fields, the Dzialoshinskii
Moriya ladder appears to behave similarly to a regular spin ladder
albeit with renormalized gaps.

\subsection{Behavior in small magnetic field}
At small $h$, we can expand equation (\ref{pol3or}) in powers of $h^2$
and find a solution valid to first order in $h^2$.  The equation
expanded to first order in $h^2$ reads:
\begin{equation}
\label{eqexp}
\prod_{i=1,2,3}(\epsilon^2-\epsilon_i^2)-2h^2 \epsilon^4 + 2h^2 A({\bf m},\alpha)
\epsilon^2 +2h^2 C({\bf m},k,\alpha)=0.
\end{equation}
One can search for solutions of the form:
$\epsilon^2(k)=\epsilon_i^2(k)+\beta_i h^2$.
We find:
\begin{equation}
\beta_i=-\frac{\lbrack
C(k,{\mathbf{m}},\alpha)+A({\mathbf{m}},\alpha)\epsilon_i(0)^2-\epsilon_i(0)^4\rbrack}
{\epsilon_i(0)\Pi_{i\ne j}(\epsilon_i(0)^2-\epsilon_{j}(0)^2)}.
\end{equation}

 From the knowledge of the spectrum we evaluate the total
magnetization through the following relation:

\begin{equation}
\label{magna}
M=\sum_{i=1}^3\sum_{\epsilon_i(k)<0} \frac{\partial \epsilon_i(h)}{\partial
h}=-\sum_{i=1}^3 \sum_{\epsilon_i(k)<0}
\frac{h}{\epsilon_i(0)}\frac{\lbrack -\epsilon^4_i(0)+
A^2({\mathbf{m}},\alpha)\epsilon^2_i(0)+C(k,{\mathbf{m}},\alpha)\rbrack}
{\Pi_{i\ne j}(\epsilon_i^2(0)-\epsilon_j^2(0))}+O(h^3).
\end{equation}

\noindent We can perform analytically the integral and the result
gives a finite magnetic susceptibility in zero magnetic field:

\begin{equation}
\chi(0)\simeq \frac{\lbrack m^2m_3^2 \ln(\frac{m}{m_3}) +(m^2+m^2m_4^2-m_4^2)
\ln(\frac{m}{m_4})+(m_3^2+m_3^2m_4^2-m_4^2)
\ln(\frac{m_3}{m_4})\rbrack}{(m^2-m_3^2)(m^2-m_4^2)(m_3^2-m_4^2)}.
\end{equation}

Thus, at zero temperature, there is a finite susceptibility in the
direction perpendicular to the DM vector, but zero susceptibility
parallel to the DM vector.

In presence of the DM+KSEA interaction the equation (\ref{eqexp}) still holds with the
prescription $\epsilon_1=\epsilon_2$ and $m=m_3$, thus we search for solutions of the form:
  $\epsilon^2(k)=\epsilon_1^2(k)+\alpha_1h+\alpha_2h^2$ and
  $\epsilon^2(k)=\epsilon_3^2(k)+\alpha_3 h^2$. We find:
\begin{eqnarray}
& &\alpha_1=\sqrt{\frac{\lbrack
C(k,{\mathbf{m}},\alpha)+A({\mathbf{m}},\alpha)\epsilon_1(0)^2-\epsilon_1(0)^4\rbrack}
{(\epsilon_1(0)^2-\epsilon_{3}(0)^2)}}; \alpha_2=0 \nonumber \\
& &\alpha_2=\sqrt{\frac{\lbrack
B(k,{\mathbf{m'}},\alpha)-\epsilon_1(0)^2\rbrack}
{(\epsilon_1(0)^2-\epsilon_{3}(0)^2)}};\alpha_1=0 \nonumber \\
& & \alpha_3=-\frac{\lbrack
C(k,{\mathbf{m}},\alpha)+A({\mathbf{m}},\alpha)\epsilon_3(0)^2-\epsilon_3(0)^4\rbrack}
{\epsilon_3(0)(\epsilon_3(0)^2-\epsilon_{1}(0)^2)^2}.
\end{eqnarray}

The first solution gives zero contribution to the net magnetization,
thus we find:

\begin{equation}
M=\sum_{\epsilon_{1,3}(k)<0} h\epsilon_1(0)
\sqrt{\frac{\lbrack
B(k,{\mathbf{m'}},\alpha)-\epsilon_1(0)^2\rbrack}
{(\epsilon_1(0)^2-\epsilon_{3}(0)^2)}}-
\frac{h}{\epsilon_3(0)}\frac{\lbrack -\epsilon^4_3(0)+
A^2({\mathbf{m}},\alpha)\epsilon^2_3(0)+C(k,{\mathbf{m}},\alpha)\rbrack}
{(\epsilon_3^2(0)-\epsilon_1^2(0))^2}+O(h^3).
\end{equation}

The integration of such expression gives the zero temperature
susceptibility in presence of the DM+KSEA interaction.

%\bibliographystyle{prsty}
%\bibliography{totphys,suthreelast,spinchiral2,dm}

\end{document}